\documentclass[conference]{IEEEtran}
\IEEEoverridecommandlockouts

\pdfoutput=1
\usepackage{cite}
\usepackage{amsmath,amssymb,amsfonts}
\usepackage{algorithmic}
\usepackage[algo2e,ruled,linesnumbered]{algorithm2e}
\usepackage{graphicx}
\usepackage{graphics}
\usepackage{epsfig}
\usepackage{epstopdf}
\usepackage{textcomp}
\usepackage{xcolor}
\usepackage[table]{xcolor}
\usepackage{array}
\usepackage{multirow}
\usepackage{booktabs}
\usepackage{float}
\usepackage{setspace}
\usepackage{color}
\usepackage{url}
\usepackage{verbatim}
\usepackage{stfloats}
\usepackage{balance}

\usepackage{chemformula}

\usepackage{subfigure}
\usepackage[caption=false,font=normalsize,labelfont=sf,textfont=sf]{subfig}

\usepackage{titlesec}

\titlespacing*{\section}
{0pt}
{0.5ex}
{0.5ex}

\titlespacing*{\subsection}
{0pt}
{0.4ex}
{0.3ex}

\titlespacing*{\subsubsection}
{0pt}
{0.3ex}
{0.2ex}
\def\BibTeX{{\rm B\kern-.05em{\sc i\kern-.025em b}\kern-.08em
T\kern-.1667em\lower.7ex\hbox{E}\kern-.125emX}}
\begin{document}

\title{Coordinated Offloading: Cooperative Multi-Agent Reinforcement Learning for the Edge--Cloud Continuum}

\author{
\IEEEauthorblockA{
Muhammad Rafid$^{\dag\S}$,
Golshan Famitafreshi$^{\ddag}$,
Vasilis Avgerinos$^{\dag}$,
Kostas Ramantas$^{\ddag}$,
Christos Verikoukis$^{\dag*}$
}
\IEEEauthorblockA{
\textit{$^{\dag}$ ISI/ATHENA RC, 26504 Patra, Greece (e-mail: \{mrafid, avgerinos, cveri\}@isi.gr)}\\
\textit{$^{\S}$ Universitat Polit\`ecnica de Catalunya (UPC), Barcelona, Spain; $^{*}$ University of Patras, 26504 Patra, Greece}\\
\textit{$^{\ddag}$ Iquadrat Informatica S. L., Barcelona, Spain (e-mail: \{g.famitafreshi, kramantas\}@iquadrat.com)}
}
\vspace{-4mm}}
\maketitle

\begin{abstract}
The increasing demand for intelligent, low-latency services in edge--cloud continuum systems poses new challenges for dynamic and efficient task offloading. We propose a Multi-Agent Reinforcement Learning (MARL) framework for distributed task offloading under partial observability, where each device \emph{offloads only a portion of its computation} to a target edge server while \emph{executing the remaining portion locally} (partial offloading).  The problem is formulated as a Markov Decision Process (MDP), where each device independently learns an optimal policy under partially observable system states. To address the non-convexity, high-dimensional state space, and continuous action space, we develop a multi-agent Advantage-Weighted Actor-Critic (AWAC)-enhanced Proximal Policy Optimization (PPO)-based computation offloading algorithm. The proposed method combines the stable policy updates of PPO with an advantage-weighting mechanism inspired by AWAC to emphasize higher-quality actions during training. To encourage cooperation without inter-device communication, a shared global reward aligned with system cost is employed. Extensive simulations demonstrate rapid convergence and significant improvement in overall system cost compared with baselines, e.g., Deep Deterministic Policy Gradient (DDPG) and Deep Q-Network (DQN).
\end{abstract}

\section{Introduction}
In recent years, the emergence of large-scale intelligent Mobile Devices (MDs) has led to the development of applications such as online gaming, VR/AR, and autonomous driving \cite{khan2019edge}. These applications are both latency critical to satisfy user experience requirements and computationally intensive. To address these challenges, Mobile Edge Computing (MEC) is the extension of cloud computing that can provide MDs with both satisfactory computing resources and task latency by deploying Edge Servers (ESs) in close proximity to MDs \cite{pham2020survey}. One of the key issues of MEC is optimal task offloading, which decides whether, where, and how much an MD offloads the task in each time slot \cite{dong2024task}.

Most existing studies have focused on MEC task offloading by assuming resource availability at ESs, and have formulated the offloading strategy as a Mathematical Programming (MP) problem to obtain closed-form or near-optimal solutions \cite{ren2022efficient}. However, these MP based approaches become computationally prohibitive in large-scale, time-varying networks. In such settings, system states, channel conditions, and task arrivals change rapidly across time slots, leading to high-dimensional offloading variables\cite{djigal2022machine}. Moreover, they generally neglect partial observability and the need for online adaptation, making them unsuitable for dynamic multi-edge-cloud settings. To address these limitations, recent works have adopted Deep Reinforcement Learning (DRL) to model the multi-slot offloading process as a MDP and solve it using RL by setting cost minimization as the reward~\cite{shakarami2020survey}. Our study focuses on partial offloading across multiple time slots and proposes a Multi-Agent Reinforcement Learning (MARL) framework that enables distributed decision making under dynamic network conditions while maintaining coordination among agents.

A large variety of studies have focused on developing RL-based offloading strategies in MEC systems~\cite{mustafa2023reinforcement}. However, most rely on centralized control, where a global controller collects system states (e.g., task size, energy level, channel condition) from all MDs and distributes offloading decisions (e.g., target ES and offloading ratio) to each device~\cite{huang2019deep}. In large-scale networks, such centralized schemes impose excessive communication overhead and become impractical due to the heavy burden of collecting all MDs’ states for each decision-making cycle~\cite{canese2021multi}. To overcome these limitations, recent works have shown increasing interest in decentralized offloading (dec\mbox{-}offloading)~\cite{lin2024decentralized}, mainly implemented through decentralized MARL frameworks, including Multi-Agent Deep Deterministic Policy Gradient (MADDPG)-based methods \cite{guo2022multi}, Counterfactual Multi-Agent (COMA)-based methods \cite{liu2020distributed}, and Multi-Agent Twin Delayed Deep Deterministic Policy Gradient (MATD3)-based methods \cite{zhao2022multi}. In these approaches, each MD acts as an independent agent and makes decisions based only on its local observations. This paradigm enables scalable and adaptive offloading under partial observability. While some works \cite{gao2023ddpg} assume inter-MD communication to aid exploration and collaboration, such signaling is often impractical when  Device-to-Device (D2D) links are unavailable. Consequently, most studies \cite{liu2020distributed} adopt an actor–critic approach with Centralized Training and Decentralized Execution (CTDE). In CTDE, a centralized critic has access to global states during training to guide learning, while at execution each actor (per-device policy) uses only its local observation.

Following this paradigm, we employ an AWAC-enhanced PPO-based policy in which the actor outputs a continuous offloading ratio $\rho \in [0,1]$ and a target ES selection $\delta$.
\begin{figure}[t]
\centering
{{\includegraphics[width=\linewidth, height=6.4cm,]{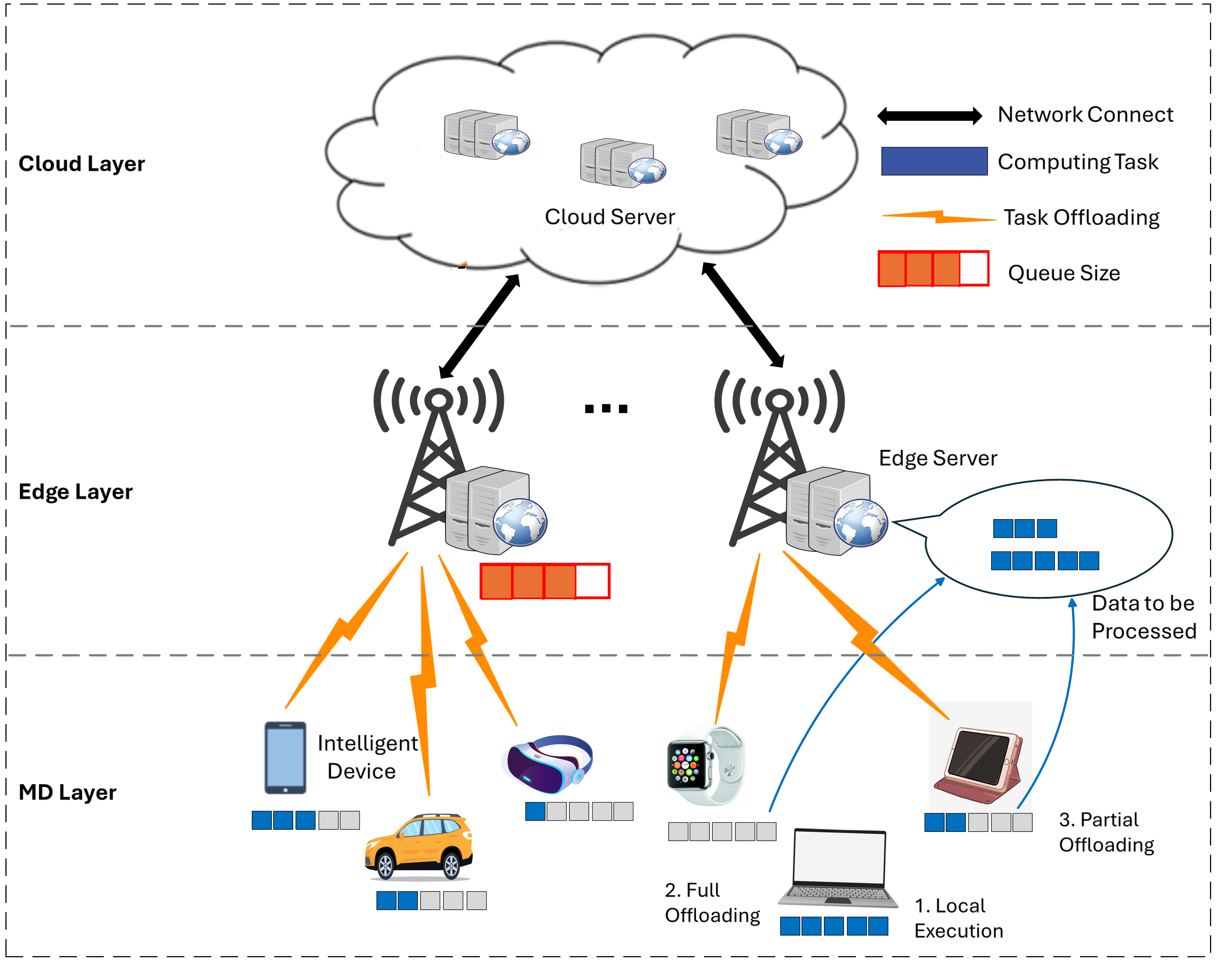}}}
  \caption{System model of the edge--cloud continuum architecture, consisting of multiple MDs, ESs, and a main controller. Each MD performs partial offloading of computation tasks to nearby ESs, while the main controller coordinates global learning and system management.}
  \label{spectrum diagram}
\end{figure}
In this paper, we consider a multi--edge--cloud continuum in which each MD determines the optimal offloading ratio and target ES under time-varying network conditions. Unlike DQN \cite{li2018task} based algorithms designed for discrete action spaces, we develop an AWAC-enhanced PPO-based computation offloading method that effectively supports a continuous action space. Each agent autonomously makes decisions from its local observations of the environment state, while a shared global reward encourages cooperative behavior guiding agents to minimize the overall system cost rather than optimizing only individual performance. \\
The main contributions of this paper can be summarized as follows.
\begin{itemize}
  \item Considering time-varying channels and slot-based operation in a multi--edge--cloud MEC system, we formulate \emph{partial offloading} (how much to offload and which ES to select) as an MDP to minimize the long-term system cost.
  \item Owing to high-dimensional state dynamics and the need for a \emph{continuous} offloading ratio, we develop an \emph{AWAC-enhanced PPO}-based computation offloading approach under the CTDE paradigm, where each MD acts as an autonomous agent using local observations at execution time, while the critic exploits global information during training.
  \item Performing extensive simulations across diverse channel and load conditions show that the proposed method achieves \emph{lower overall system cost} and \emph{stable convergence} compared with other baselines.
\end{itemize}

The rest of this paper is organized as follows. Section II presents the system model of the multi--edge--cloud continuum, including the communication, computation, and system cost models. Section III presents the partial offloading optimization problem and the proposed MARL framework. Section IV provides performance evaluation results and comparison with representative baseline schemes. Finally, Section V concludes the paper and discusses future research directions.
\section{System Model}
We consider a multi–edge–cloud continuum architecture, as shown in Fig. \ref{spectrum diagram}, consisting of three layers: the device  layer, the edge layer, and the cloud layer. Let $\mathcal{S}=\{1,\dots,S\}$ and $\mathcal{U}=\{1,\dots,U\}$ denote the sets of ESs and MDs, respectively, and let $\mathcal{N}=\{1,\dots,N\}$ be the set of time slots. MDs are randomly distributed across Base Stations (BSs), and the system time is equally divided into slots indexed by $n\in\mathcal{N}$. Each ES $s\in\mathcal{S}$ has computing capacity $f_s$ and each ES $s$ maintains a task queue $q_s$, where $q_s^{n}$ denotes the set of tasks that must be executed before the current task on ES $s$ (i.e., the queue size) at slot $n$.
Each MD $u\in\mathcal{U}$ is equipped with a local CPU of capacity $f_u$ and an energy budget $E_u^{\max}$, with $E_u^{n}$ denoting the available energy at slot $n$. At every slot $n$, MD $u$ generates a computation-intensive and latency-sensitive task $J_u^{n}=(C,L_u^{n},D^{\max})$, where $C$ is the number of CPU cycles required, $L_u^{n}$ is the input task size, and $D^{\max}$ is the maximum tolerable delay. Each MD maintains an offloader $\pi_u$ \footnote{Similar to \cite{zhan2020deep}, we consider partial offloading, which can be readily extended to binary offloading by constraining the offloading ratio to either 0 or 1.} that partially offloads $J_u^{n}$ to a target ES each slot. The offloading decision is
 \begin{equation*}
a_u^{n}=\{\rho_u^{n},\,\delta_u^{n}\},
\end{equation*}
\begin{equation}
\rho_u^{n}\in[0,1].
\label{eq:gain-1}
\end{equation}
\begin{equation}
\delta_u^{n}\in\mathcal{S},\ \text{if }\rho_u^{n}\neq0.
\label{eq:gain-2}
\end{equation} where $\rho_u^{n}$ denotes the offloaded fraction and $\delta_u^{n}$ denotes the selected ES. This offloading model can be applied to a wide range of application scenarios, such as malware detection \cite{burguera2011crowdroid}. Malware detection systems analyze the runtime behavior of a large number of applications running on MDs by logging execution data for each application. These application traces must be scanned in real time using the latest malware signature files obtained from a security database. By offloading detection tasks to secure servers, edge-based malware detection can significantly reduce the computational burden on MDs. In this scenario, users may offload any portion of the traces to the edge server, where detection algorithms are executed to ensure system security \cite{zhan2020deep}. Due to the absence of direct D2D communications in most practical scenarios, the offloading controllers operate independently but collectively aim to minimize the long-term system cost.
\subsection{Communication Model}
Following the standard block fading channel model \cite{meng2019power}, at slot $n$, the uplink channel gain between MD $u$ and ES $s$ over the offloading link $\mathcal{L}^{u\to s}_{n}$ is modeled as
\begin{equation}
h_{u,s}^{n}=\alpha_{u,s}^{n}\,\bigl|g_{u,s}^{\,n}\bigr|^{2},
\label{eq:gain-usn}
\end{equation}
where $\alpha_{u,s}^{n}$ accounts for large–scale effects (path loss and shadowing) and $g_{u,s}^{\,n}$ is the small–scale fading term. The large–scale component with log–normal shadowing is written as
\begin{equation}
{\alpha}_{u,s}^{n} = -148.1 - 37.6\,\log_{10}\!\bigl(d_{u,s}^{n}\bigr) + 10\log_{10} z,
\label{eq:PL-db}
\end{equation}
where $d_{u,s}^{n}$ is the link $\mathcal{L}^{u\to s}_{n}$ distance and $z$ is a log–normal random variable. The small–scale fading follows the Jakes model (first–order complex Gaussian Markov process).

Assuming multiple MD–ES offloading links coexist in slot $n$, the aggregate uplink interference measured at ES $s$ (excluding user $u$) is
\begin{equation}
I_{s}^{n}=\sum_{\substack{u'\in\mathcal{U}\\ u'\neq u}}
\mathbf{1}\!\bigl(\rho_{u'}^{n}\neq 0\bigr)\,
p^{\mathrm{tx}}\, h_{u',s}^{n},
\label{eq:interference}
\end{equation}
where $\rho_{u'}^{n}$ is the offloading ratio of device $u'$, $p^{\mathrm{tx}}$ its transmit power, and $\mathbf{1}(\cdot)$ is the indicator function. Hence, the transmission rate achievable for the link $\mathcal{L}^{u\to s}_{n}$ in the slot $n$ is,
\begin{equation}
R_{u,s}^{n}
= B\log_{2}\!\left(1+\frac{p^{\mathrm{tx}}\,h_{u,s}^{n}}{I_{s}^{n}+\sigma^{2}}\right),
\label{eq:rate}
\end{equation}
where $B$ denotes the bandwidth and $\sigma^{2}$ the Gaussian noise power.
\subsection{Local Computing Model}
 The partial offloading strategy is applied to MD $u$’s task $J_u^{n}$ at slot $n$. If $\rho_u^{n}\neq 1$, the remaining fraction $(1-\rho_u^{n})$ of $J_u^{n}$ is executed locally. Accordingly, the local task latency and energy consumption of MD $u$ are,
\begin{equation}
t_{u}^{\text{loc},n}=\frac{(1-\rho_{u}^{n})\,CL_u^{n}}{f_{u}}, 
\qquad
e_{u}^{\text{loc},n}=\xi\,f_{u}^{2}\,(1-\rho_{u}^{n})\,CL_u^{n},
\label{eq:local-lat-energy}
\end{equation}
where $f_{u}$ is the device CPU frequency and $\xi$ is the energy coefficient of the MD’s chip.
\subsection{Edge Computing Model}

If $\rho_u^{n}\neq0$, the fraction $\rho_u^{n}$ of $J_u^{n}$ is offloaded to the selected ES $s$ for remote execution in slot $n$. The end-to-end latency comprises three parts: (i) the \emph{transmission time} to deliver $\rho_u^{n}L_u$ bits over $\mathcal{L}_{n}^{u\to s}$, given by $t_{u,s}^{\mathrm{tx},n}=\rho_u^{n}L_u^{n}/R_{u,s}^{n}$; (ii) the \emph{waiting time} at ES $s$ due to the queued workload and the offloaded tasks that complete transmission to $s$ earlier in the same slot, $t_{u,s}^{\mathrm{wait},n}=\bigl(q_s^{n}+\sum_{u'\in\mathcal{U}'} \rho_{u'}^{n} L^n_{u'}\bigr)C/f_s$, where $\mathcal{U}'=\{\,u'\in\mathcal{U}\mid u'\neq u,\ \rho_{u'}^{n}\neq 0,\ \delta_{u'}^{n}=s,\ t_{u',s}^{\mathrm{tx},n}<t_{u,s}^{\mathrm{tx},n}\,\}$ collects the MDs that finish transmission to $s$ before $u$ in slot $n$; and (iii) the \emph{execution time} of the offloaded portion on the ES CPU, $t_{u,s}^{\mathrm{exe},n}=\rho_u^{n}L_u^{n}C/f_s$. Accordingly, the edge-side delay and the MD’s offloading energy are
\begin{equation}
t^{\mathrm{edge},n}_{u,s}
= t_{u,s}^{\mathrm{tx},n}+t_{u,s}^{\mathrm{wait},n}+t_{u,s}^{\mathrm{exe},n},
\label{eq:edge-latency-final}
\end{equation}
\begin{equation}
e^{\mathrm{edge},n}_{u,s}
= p_{u}^{\mathrm{tx},n}\,t_{u,s}^{\mathrm{tx},n},
\label{eq:edge-energy-final}
\end{equation}

\subsection{ System Cost Model}
The considered system cost model includes two main parts namely time and energy which are explained as follows. 

\emph{1) Time Cost:} Since $J_u^{n}$ can be executed simultaneously on MD $u$ ($\rho_u^{n}\neq1$) and on ES $s$ ($\rho_u^{n}\neq0$), the task latency of $J_u^{n}$ is determined by the longer execution time on the two sides, which can be written as
\begin{equation}
t_u^{n}=\max\!\left(t_u^{\mathrm{loc},n},\ t_{u,s}^{\mathrm{edge},n}\right),
\end{equation}
and it should satisfy the delay constraint,
\begin{equation}
t_u^{n}\le D^{\max}.
\label{max}
\end{equation}

\emph{2) Energy Cost:} Different from the time cost, the total energy consumed for task $J_u^{n}$ includes both local computing and edge offloading, which is given by
\begin{equation}
e_u^{n}=e_{u,s}^{\mathrm{edge},n}+e_u^{\mathrm{loc},n}.
\label{ent}
\end{equation}
For each slot $n$, the remaining energy budget of MD $u$ is
\begin{equation}
E_u^{n}=E_u^{\max}-e_u^{n}, \quad E_u^{n}\ge0.
\label{en}
\end{equation}

\emph{3) System Cost:} The overall cost for executing task $J_u^{n}$ is defined as a weighted sum of latency and energy consumption, $c_u^{n}=\eta\,t_u^{n}+(1-\eta)\,e_u^{n}$, where $\eta\in[0,1]$ balances delay and energy. The average cost per slot is $\Psi^{n}=\tfrac{1}{U}\sum_{u=1}^{U}c_u^{n}$, 
and the average system cost over all $N$ slots is calculated as
\begin{equation}
\Psi=\tfrac{1}{N}\sum_{n=1}^{N}\Psi^{n}.
\label{cost}
\end{equation}

\section{Problem Formulation and Proposed MARL Framework}
\subsection{Optimization Problem Formulation}
Based on the model introduced above, we formulate the optimization problem for our edge–cloud continuum. Since unreasonable offloading strategies may waste computational resources, we explicitly optimize the offloading ratio and the target ES. Viewing the offloading ratio $\rho_u^{n}$ and the destination $\delta_u^{n}$ as the optimizing variables, $c$ can be considered as a function of $\rho_u^{n}$ and $\delta_u^{n}$, i.e., $c\!\big(\rho_u^{n},\delta_u^{n}\big)$.
Our goal is to keep the average system cost as low as possible over the whole episode.
\begin{equation}
\mathcal{P}:\quad 
\min_{\rho_u^{n},\,\delta_u^{n}}\, c\!\left(\rho_u^{n},\delta_u^{n}\right),
\label{pro}
\end{equation}
\begin{equation*}
 \text{s.t. Eqs.~}\eqref{eq:gain-1}, \eqref{eq:gain-2}, \eqref{max}, \text{ and } \eqref{en}.
\end{equation*}

The above problem is challenging to solve since it has \textit{ NP-hard} nature. The main idea is to draw an analogy between the considered problem and the Generalized Assignment Problem (Section 15.8.5 in \cite{hooker2006operations}). In essence, it reduces to assigning $M$ tasks to $S$ servers such that the total load assigned to each server does not exceed its available capacity while the objective is minimized.

According to Eq. \eqref{cost}, minimizing $c$ in Eq. \eqref{pro} is equivalent to minimizing the accumulated per-slot costs across $N$ slots. Because the cost is inversely related to the reward, $\mathcal{P}$ is aligned with a discrete-time MDP that aims to maximize the cumulative reward over multiple steps.
\subsection{AWAC-Enhanced PPO-Based MARL Framework}
Concerning the heavy burden of collecting all MDs' states at every slot and the poor scalability of centralized scheduling, we adopt a decentralized MARL formulation. To enable distributed decision-making under dynamic and partially observable network conditions, the proposed offloading framework is modeled as an MDP, where each MD operates as an autonomous learning agent interacting with its environment. Traditional DQN \cite{li2018task} algorithms are limited by their discrete action representation and are not suited for continuous control in resource allocation and task offloading. Accordingly, we adopt an \emph{AWAC-enhanced PPO-based MARL approach} to learn adaptive policies that minimize the long-term system cost \cite{chandler2025network}.

Following the standard definition of MDP \cite{shakarami2020survey} $(\mathcal{S},\mathcal{A},\mathcal{P},\mathcal{R})$, each agent observes its local state $s_u^n \in \mathcal{S}$ at slot $n$, selects an action $a_u^n \in \mathcal{A}$ through its policy $\pi_u(a_u^n \mid s_u^n)$, and receives an instantaneous reward $r_u^n$. The transition probability $\mathcal{P}(s^{n+1}\!\mid s^n,a^n)$ captures the temporal evolution of network dynamics, including channel variations, queue status, and energy availability. Under CTDE, a centralized critic estimates the state-value function \(V(s_u^n)\), while each actor executes decisions using only local observations. The clipped advantage is defined as
\begin{equation}
A_u^n = \mathrm{clip}(R_u^n - V(s_u^n), -c, c),
\label{eq:adv}
\end{equation}
where \(R_u^n\) denotes the empirical return of agent \(u\) at slot \(n\), and \(c\) is a clipping threshold introduced for numerical stability. To prioritize high-quality actions, an exponential advantage weight is introduced as
\begin{equation}
w_u^n = \exp(A_u^n/\tau), \qquad
w_u^n \leftarrow \min(w_u^n, w_{\max}),
\label{eq:weight}
\end{equation}
where \(\tau\) controls the weighting sharpness and \(w_{\max}\) limits excessively large updates. The actor is then optimized using the weighted PPO clipped surrogate objective
\begin{equation}
\small
L_{\mathrm{AW\text{-}PPO}}
=
\mathbb{E}_{u,n} \left[
w_u^n
\min\!\left(
r_u^n(\theta)\hat{A}_u^n,\;
\mathrm{clip}\!\left(r_u^n(\theta), 1-\epsilon, 1+\epsilon\right)\hat{A}_u^n
\right)
\right]
\label{eq:awppo}
\end{equation}
where
\begin{equation}
r_u^n(\theta)
=
\frac{\pi_{\theta}(a_u^n \mid s_u^n)}
{\pi_{\theta_{\mathrm{old}}}(a_u^n \mid s_u^n)}.
\label{eq:ratio}
\end{equation}
The total training objective is given by
\begin{equation}
L_{\mathrm{total}} = L_{\mathrm{AW\text{-}PPO}} + c_1 L_{\mathrm{value}} - c_2 H[\pi_{\theta}],
\label{eq:total}
\end{equation}
where \(L_{\mathrm{value}}\) is the value loss and \(H[\pi_{\theta}]\) is the entropy regularization term. 
 This formulation combines PPO's stable policy updates with AWAC-style advantage weighting, thereby improving learning efficiency and convergence under dynamic edge--cloud conditions.
The reward plays a crucial role in stabilizing the learning process and aligning individual behaviors toward the global objective. In this work, the episodic reward of each agent is defined as an inverse cost function that combines local performance with a lightweight global cooperative component. For each agent $u$, the episodic reward is 
\begin{equation}
\small
r_u =
\begin{cases}
\dfrac{1}{\bar{c}_u}+\lambda_{\text{team}}\dfrac{1}{\bar{c}_{\text{global}}},
& \text{if } (V_t=0,\; V_e=0), \\[1pt]
-\big(\beta_t V_t+\beta_e V_e\big)+\lambda_{\text{team}}\dfrac{1}{\bar{c}_{\text{global}}},
& \text{otherwise}.
\end{cases}
\end{equation}
where $\bar{c}_u$ is the average cost of the agent’s episode, $\bar{c}_{\text{global}}$ is the mean cost for all agents, and $V_t, V_e$ denote cumulative delay and energy budget violations, respectively. The coefficients $\beta_t,\beta_e>0$ and $\lambda_{\text{team}}\!\ge\!0$ balance penalties and cooperation. A \emph{positive local reward} is granted only when the episode has no violations, while any violation yields a \emph{negative penalty} proportional to the degree of constraint violation. The global cooperative term $\lambda_{\text{team}}/\bar{c}_{\text{global}}$ is included in \emph{both} cases to provide a shared convergence signal and promote system-wide coordination.

Through iterative interaction with the environment, each agent updates its policy using the proposed AWAC-enhanced PPO-based actor-critic framework. The centralized critic estimates state values and provides advantage-guided learning signals, while the actor is optimized using a weighted PPO clipped objective. In this way, the agents collectively minimize the long-term system cost and achieve stable and cooperative convergence in distributed operation.

\section{Performance Evaluation}
We consider an edge--cloud continuum deployed over a $0.5\times 0.5~\text{km}^2$ area, where $10 \sim 15$ MDs are distributed per MEC node. The trajectories of the MDs are generated by a random walk mobility model across multiple MEC nodes. Each MD is associated with a computation--intensive and latency--sensitive task that is partially offloaded to a target MEC server for execution. For task $J_u^n$, the input data size $L_u^n$ is sampled uniformly from $\mathcal{U}(0.8,\,1.2)$ MB. The required CPU cycles per bit are set to $s = 1000$~cycles/bit. The system bandwidth is set to $B=1~\text{MHz}$. The receiver noise power is $\sigma^2=-100~\text{dBm}$~\cite{wang2021computation}, and the MD uplink transmit power is $P_{\text{tx}}=0.1~\text{W}$. Each MD has a maximum CPU frequency $f_u\in[0.6,\,1]~\text{GHz}$. The required MEC edge servers are equipped with CPUs of $f_s\in[4,\,8]~\text{GHz}$. Unless otherwise stated, the maximum tolerable task delay is $D_{\max}\in[0.6,\,1]~\text{s}$. The numerical simulation of the MEC system is implemented in Python, and the primary simulation parameters are listed in Table \ref{tab:primary_params}. For the proposed AWAC-enhanced PPO\mbox{-}based offloading framework, the discount factor is set to $\gamma = 0.99$. We use a replay buffer of size $20{,}000$ and a mini\mbox{-}batch size of $256$ for training, following the implementation settings in \cite{chandler2025network}.
\begin{table}[t]
\centering
\caption{MEC Experiment Parameters.}
\label{tab:primary_params}
\begin{tabular}{@{}lll@{}}
\toprule
\textbf{Parameter} & \textbf{Description} & \textbf{Value} \\ 
\midrule
$S$               & Number of ESs                           & 5 \\
$A$               & Area size                               & $0.5 \times 0.5~\mathrm{km}^2$ \\
$B$               & System bandwidth                        & $1~\mathrm{MHz}$ \\
$\sigma^{2}$      & White Gaussian noise power              & $-100~\mathrm{dBm}$ \\
$P_{\mathrm{tx}}$ & MD transmit power                       & $0.1~\mathrm{W}$ \\
$f_u^{\max}$      & MD CPU                                  & [$0.6,\,1]~\mathrm{GHz}$ \\
$f_s$             & ES CPU frequency                        & $[4,\,8]~\mathrm{GHz}$ \\
$D_{\max}$        & Maximum task delay                      & [$0.6,\,$1]$~\mathrm{s}$ \\
$L_u^n$           & Task input size                         & $[0.8,\,1.2]~\mathrm{MB}$ \\
$C$               & CPU cycles/bit                          & 1000 \\
$\Delta$          & Slot length                              & $1~\mathrm{s}$ \\
$N$               & Number of slots                          & 10 \\
$\xi$             & Energy efficiency coefficient            & $10^{-27}$ \\
\bottomrule
\end{tabular}
\end{table}
For comparison, we conduct numerical experiments and evaluate the proposed algorithm against four baseline approaches:
\begin{itemize}
  \item \textbf{Local-Only:} In each time slot $n$, every MD executes the entire task locally (no offloading).
  \item \textbf{Offload-Only:} In each time slot $n$, every MD offloads the entire task to its \emph{nearest} ES for remote execution.
  \item \textbf{MA-DQN:} A multi-agent DQN \cite{li2018task} baseline with a \emph{discrete} action space, in which the offloading levels are quantized as $\rho_u^{n}=\{0,\,0.1,\,\ldots,\,1.0\}$.
  \item \textbf{MADDPG:} A MADDPG-based continuous action method, where each agent jointly learns the offloading ratio $\rho_u^{n} \in [0,1]$ and the target edge server selection $\delta_u^{n}$.
  
 \end{itemize}

We evaluate performance using three metrics: 1) \textbf{average energy consumption (AvgEngCon)}, computed from Eq. \eqref{ent} as the per-task sum of local CPU energy for the locally executed portion and uplink transmission energy for the offloaded portion, averaged over all tasks; 2) the \textbf{average task cost (AvgCost)}, defined in Eq. \eqref{cost} and reported over \emph{successful} episodes (i.e., episodes without constraint violations), which reflects the overall Quality of Service (QoS) of MEC tasks; and 3) the \textbf{task success rate (SuccRate)}, i.e., the fraction of tasks completed within the deadline during an episode
$\sum_n \mathbb{I}\left(t_n^{m, \Gamma}<t^{\max } \mid \forall m \in m\right) / N $
where $\mathbb{I}(\cdot)$ is the indicator function, $t_u^{n}$ is the realized latency of task $J_u^{n}$, and $D_{\max}$ is the per task delay bound.

\begin{figure}[!h]
\centering
\includegraphics[scale=0.40]{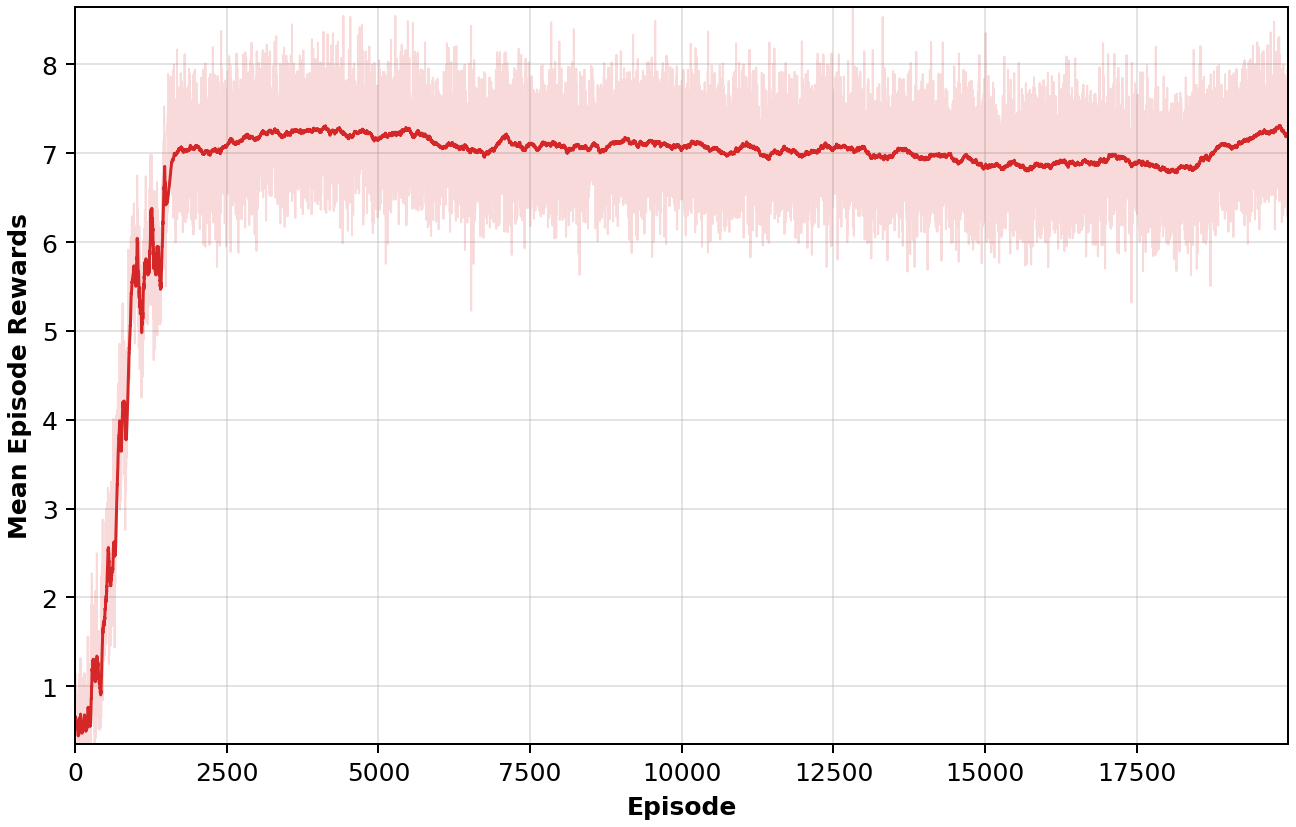}
\caption{Cumulative reward over training episodes.}
\label{fig1}
\end{figure}

As illustrated in Fig. \ref{fig1}, the reward in our model is formulated as an inverse cost function that combines each agent’s local performance with a lightweight global cooperative term. 
At the beginning of training, the cumulative reward is low and exhibits oscillations because agents explore and frequently select invalid or suboptimal actions, resulting in repeated penalties. As training progresses, the agents start to identify valid action spaces, reduce constraint violations, and achieve higher rewards. Consequently, the cumulative reward increases sharply and then stabilizes, indicating that the policy converges to a steady and optimal decision strategy.

\begin{figure}[h]
\centering
{{\includegraphics[scale=0.36]{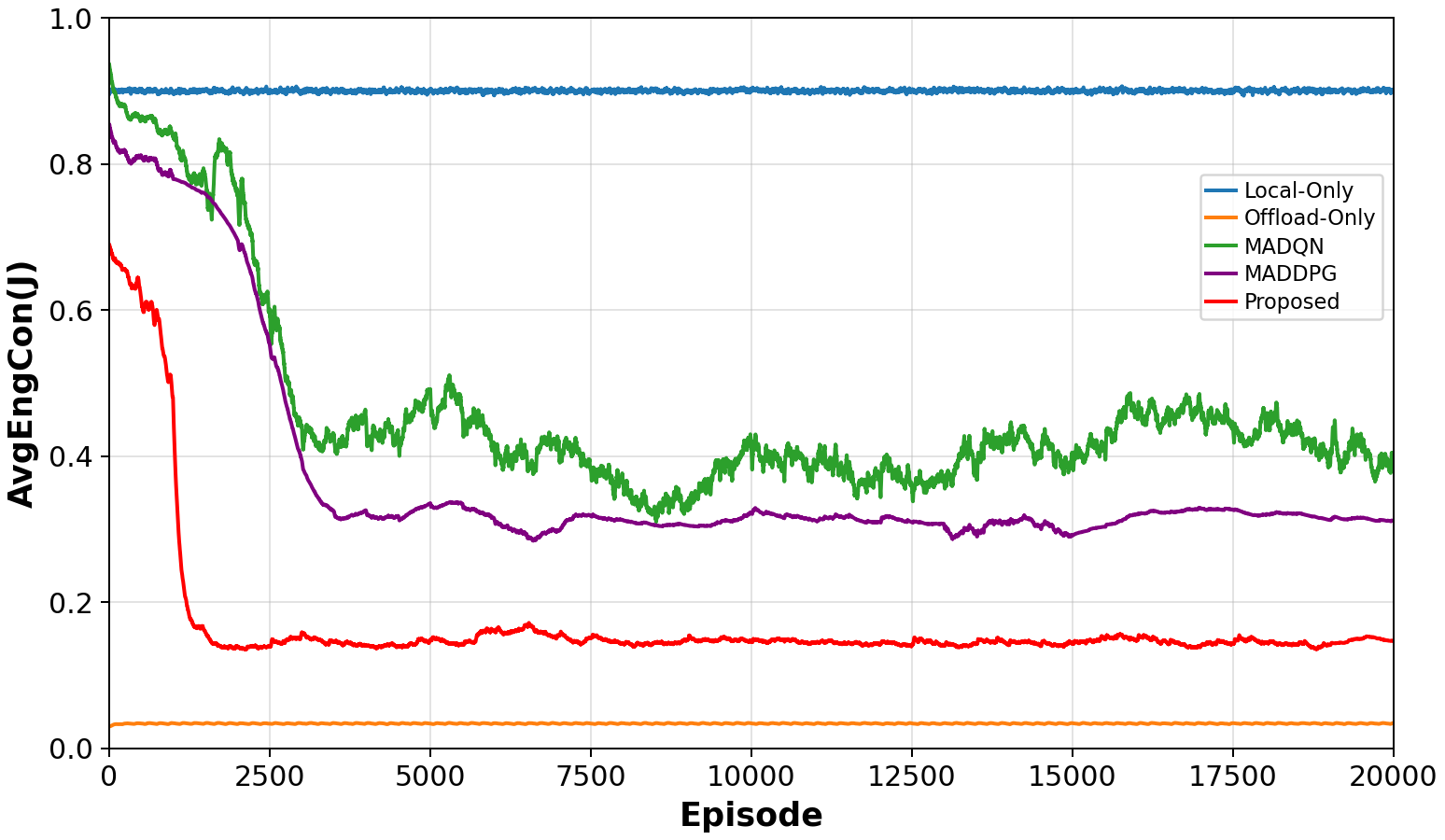}}}
 \caption{Performance comparison of different baseline offloading approaches with respect to average Energy consumption.}
  \label{fig4}
\end{figure}

Fig. \ref{fig4} illustrates the AvgEngCon trend. As observed, \textit{Local-Only} remains high and nearly flat because all computation is performed locally, whereas \textit{Offload-Only} stays very low since only the transmission energy is considered. The \textit{MA-DQN} method shows a gradual decline but stabilizes at a relatively higher level with noticeable oscillations due to its quantized offloading ratios. The \textit{MADDPG} approach achieves lower average energy consumption than MA-DQN and converges more smoothly, benefiting from its continuous action representation. In contrast, the \textit{proposed AWAC-enhanced PPO-based method} exhibits a sharp reduction during the early training stage and then converges to the lowest and most stable AvgEngCon level, indicating that it learns more energy-efficient offloading ratios \(\rho\) and edge server selections \(\delta\) under time-varying network conditions.

\begin{figure}[!h]
\centering
{{\includegraphics[scale=0.38]{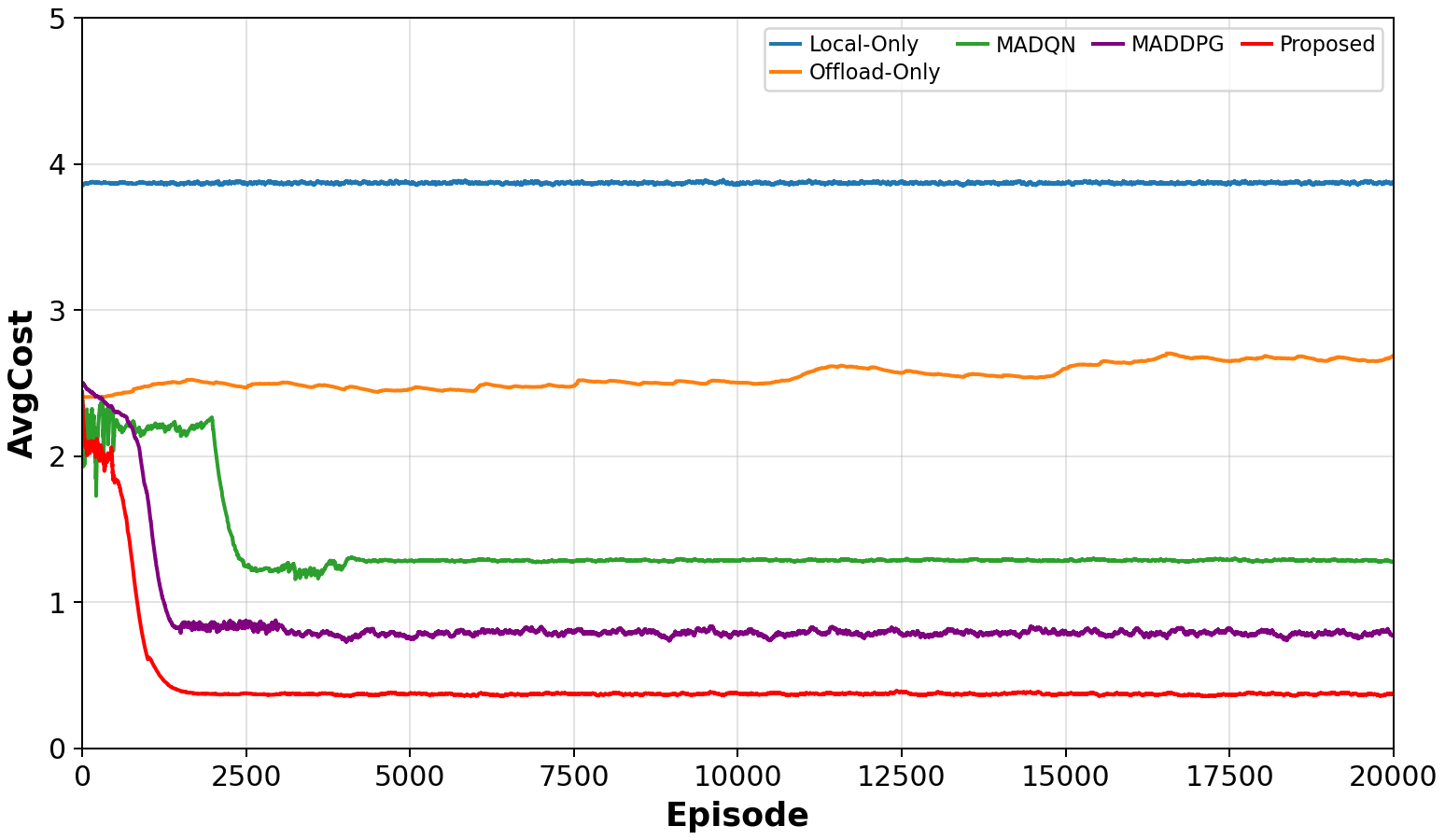}}}
 \caption{Performance comparison of different baseline offloading approaches with respect to average task cost.}
  \label{fig2}
\end{figure}
As shown in Fig. \ref{fig2}, the performance comparison of the proposed algorithm in terms of AvgCost. As observed, the \textit{proposed AWAC-enhanced PPO-based method} achieves the lowest average cost, converging rapidly to around \(0.35\). The \textit{MADDPG} approach also performs well and stabilizes at a lower cost than \textit{MA-DQN}, but remains noticeably above the proposed method. The \textit{MA-DQN} method converges more slowly and settles at a higher cost value of around \(1.25\). In contrast, the \textit{Offload-Only} strategy remains around \(2.4\text{--}2.7\), while the \textit{Local-Only} method stays near \(3.75\). Overall, the proposed method consistently outperforms all compared approaches in both convergence speed and final AvgCost. 

\begin{figure}[h]
\centering
\includegraphics[scale=0.38]{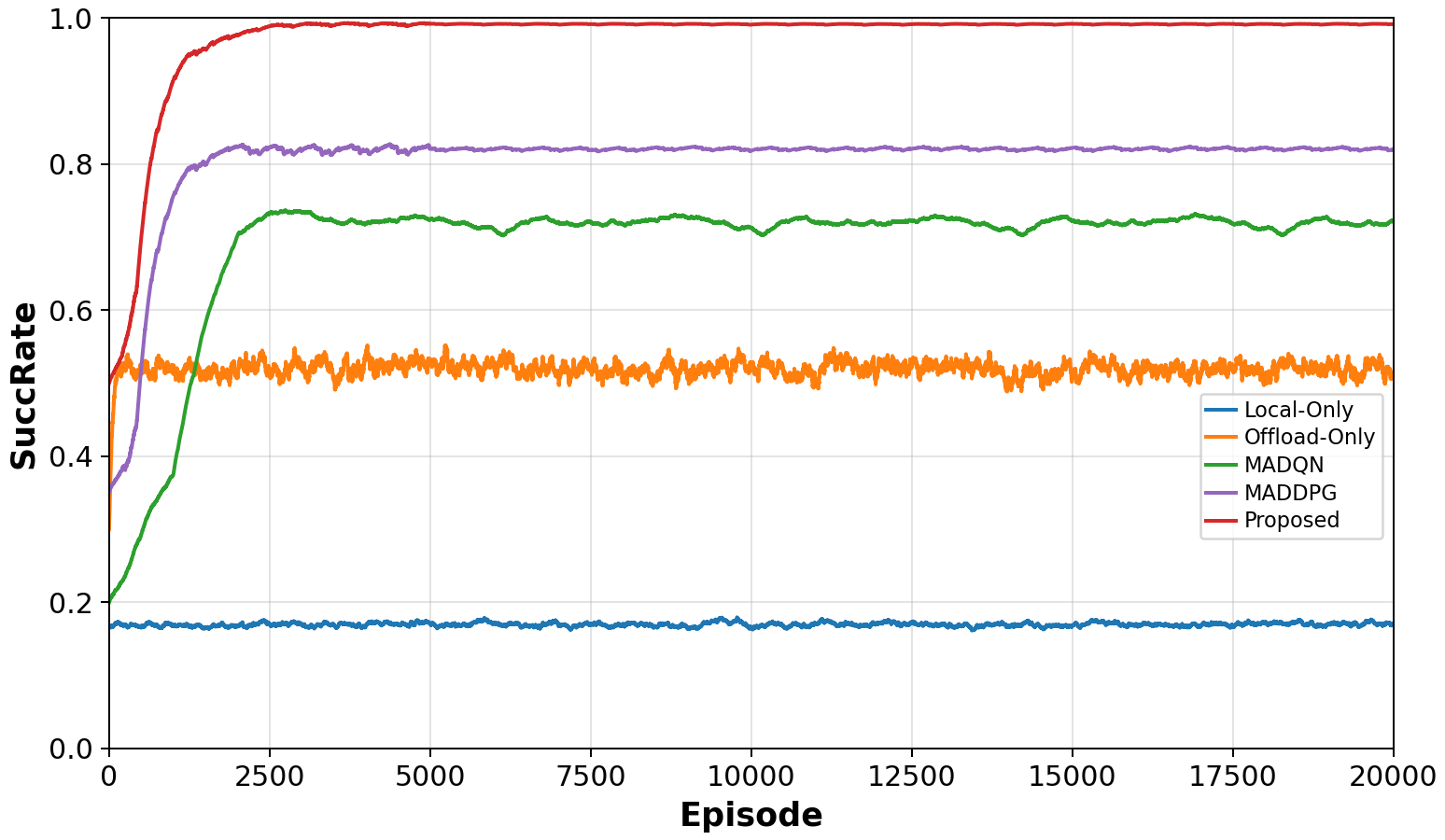}
\caption{Performance comparison of different baseline offloading approaches with respect to Success rate.}
\label{fig3}
\end{figure}

Fig.~\ref{fig3} illustrates the task SuccRate, i.e., the fraction of tasks completed within the deadline \(D_{\max}\), which serves as a reliability indicator for the MEC system. As observed, the \textit{proposed AWAC-enhanced PPO-based method} achieves the highest success rate, reaching nearly \(100\%\) after convergence, which indicates that deadline violations become very rare during stable operation. Meanwhile, the \textit{MADDPG} approach also performs strongly and stabilizes at around \(82\%\), while \textit{MA-DQN} converges more gradually and settles near \(72\%\). The \textit{Offload-Only} strategy remains around \(50\text{--}55\%\) because of persistent ES-side congestion, whereas \textit{Local-Only} stays near \(16\text{--}18\%\) under device-CPU bottlenecks. Overall, the proposed approach delivers the highest reliability and the fastest stabilization among all compared methods.
\section{Conclusion and Future Work}
This paper studied computation offloading in a multi--edge--cloud continuum, where MDs minimize the average system cost by jointly optimizing ES selection and the partial offloading ratio. To address the nonconvex and time-varying nature of the problem under partial observability and without inter-device communication, we developed a decentralized MARL framework based on an \emph{AWAC-enhanced PPO} scheme, which combines PPO's stable policy updates with AWAC-style advantage weighting and a shared global reward to guide agents toward system-wide cost reduction. We detailed the RL concepts and training pipeline, and analyzed key hyper-parameters, including the learning rate, discount factor, and training strategies. Extensive simulations show that the proposed scheme converges rapidly and achieves a lower average cost than the compared methods. These results confirm the effectiveness of stable policy optimization, advantage-guided learning, and cooperative reward design for partial offloading in dynamic edge--cloud environments. Future work
will focus on integrating a meta-guided controller to support
scalable multi-agent coordination and adaptive exploration,
with explicit optimization of communication overhead for
large-scale distributed deployments.
\section{ACKNOWLEDGMENT}
This work has received funding from the European Union under the EXACT-6G project (Grant Agreement No. 101120297), the SUNRISE-6G project (Grant Agreement No. 101139257), and the 6G-INTENSE project (Grant Agreement No. 101139266).
\bibliographystyle{IEEEtran}
\bibliography{bib}
\end{document}